\documentclass{article}

\usepackage{arxiv}

\usepackage[utf8]{inputenc} 
\usepackage[T1]{fontenc}    
\usepackage{hyperref}       
\usepackage{url}            
\usepackage{booktabs}       
\usepackage{amsfonts}       
\usepackage{nicefrac}       
\usepackage{microtype}      
\usepackage{lipsum}		
\usepackage{graphicx}
\usepackage{natbib}
\usepackage{doi}
\usepackage{graphicx,caption}
\usepackage{xcolor}
\usepackage{amsmath}
\usepackage{appendix}
\usepackage{mathtools}
\usepackage{amsmath}
\usepackage{tikz}
\usepackage{listings}
\usetikzlibrary{shapes.geometric, arrows}

\title{DaDRA: A Python Library for Data-Driven Reachability Analysis}

\date{August 2021}	

\author{\href{https://www.linkedin.com/in/jaredmejia/}{\hspace{1mm}Jared Mejia}\thanks{2021 NSF Superb Program undergraduate student participant} \\
	Department of Computer Science\\
	Pomona College\\
	Claremont, CA 91711 \\
	\texttt{jared.mejia@pomona.edu} \\
	\And
	{\hspace{1mm}Alex Devonport}\thanks{PhD student mentor for 2021 NSF Superb Program} \\
	Department of Electrical Engineering and Computer Science\\
	University of California, Berkeley\\
	Berkeley, CA 94720 \\
	\texttt{alex\_devonport@berkeley.edu} \\
	\And
	{\hspace{1mm}Murat Arcak}\thanks{Faculty advisor for 2021 NSF Superb Program} \\
	Department of Electrical Engineering and Computer Science\\
	University of California, Berkeley\\
	Berkeley, CA 94720 \\
	\texttt{arcak@berkeley.edu} \\
}

\renewcommand{\headeright}{Article}
\renewcommand{\undertitle}{Article}
\renewcommand{\shorttitle}{\textit{DaDRA} Library}

\hypersetup{
pdftitle={DaDRA Library},
pdfauthor={Jared ~Mejia, Alex, ~Devonport, Murat ~Arcak},
pdfkeywords={data-driven, reachability analysis},
}

\begin{document}
\maketitle

\begin{abstract}
	Reachability analysis is used to determine all possible states that a system acting under uncertainty may reach. It is a critical component to obtain guarantees of various safety-critical systems both for safety verification and controller synthesis. Though traditional approaches to reachability analysis provide formal guarantees of the reachable set, they involve complex algorithms that require full system information, which is impractical for use in real world settings. We present \emph{DaDRA}, a Python library that allows for data-driven reachability analysis with arbitrarily robust probabilistic guarantees. We demonstrate the practical functionality of DaDRA on various systems including: an analytically intractable chaotic system, benchmarks for systems with nonlinear dynamics, and a realistic system acting under complex disturbance signals and controlled with an intricate controller across multiple dimensions.
	
\end{abstract}

\keywords{data-driven \and reachability analysis \and safety verification \and control}

\section{Introduction}

%
%
\quad Reachability analysis is an effective method to guarantee the safety of power systems, safety-critical robots, and other nonlinear systems in the face of uncertainty.
Traditionally, approaches to reachability analysis involve complex algorithms that obtain formal guarantees of the reachable set. 
The problem with these methods is that they require full system information, but most systems of practical interest do not come in a form that is easily analyzable, as they are often high-dimensional and with imperfect information. 
\quad For data-driven reachability analysis, rather than obtaining a formal guarantee of the reachable set, data is acquired from experiments and simulations in order to estimate the reachable set with a probabilistic guarantee. 
The benefit of this approach is that virtually any system whose behavior can be simulated or measured experimentally can be evaluated with data-driven reachability analysis.

%
%
\quad Most of the current existing tools for reachability analysis employ the traditional approaches, making them impractical for analyzing real-world complex solutions.
As a solution, we propose \emph{DaDRA}, a Python library built specifically for data-driven reachability analysis.
The library allows users much of the same functionality as traditional reachability analysis tools, but the nature of the data-driven methods allows for analysis of far more complex and realistic systems. 
Furthermore, the tool provides the ability to estimate reachable sets with arbitrary desired probabilistic guarantees while taking advantage of parallelizability to accelerate the computation and allowing for insightful visualizations.
\section{Background}

%
%
\subsection{Data-Driven Reachability Analysis}
\quad In the data-driven approach to reachability analysis, we consider a dynamical system with an initial set $\mathcal{X}_0 \subseteq \mathbb{R}^{n_x}$, a set of disturbances $\mathcal{D}$, where $d: [t_0, t_1] \to \mathbb{R}^{n_x}$ and $d \in \mathcal{D}$, and a state transition function $\Phi: \mathcal{X}_0 \times \mathcal{D} \to \mathbb{R}^{n_x}$.
The goal is to estimate the image of this state transition function defined on an initial set and acted upon by disturbance signals. That is, we would like to obtain an estimate of the forward reachable set $\mathcal{R} = \{\Phi(t_1; t_0, x_0, d): x_0 \in \mathcal{X}_0, d \in \mathcal{D}\}$, including all possible evolutions of a given system in the time range $[t_0, t_1]$. 
\\
\begin{figure}[h]
    \centering
    \captionsetup{width=.8\linewidth}
    \includegraphics[width=.5\linewidth]{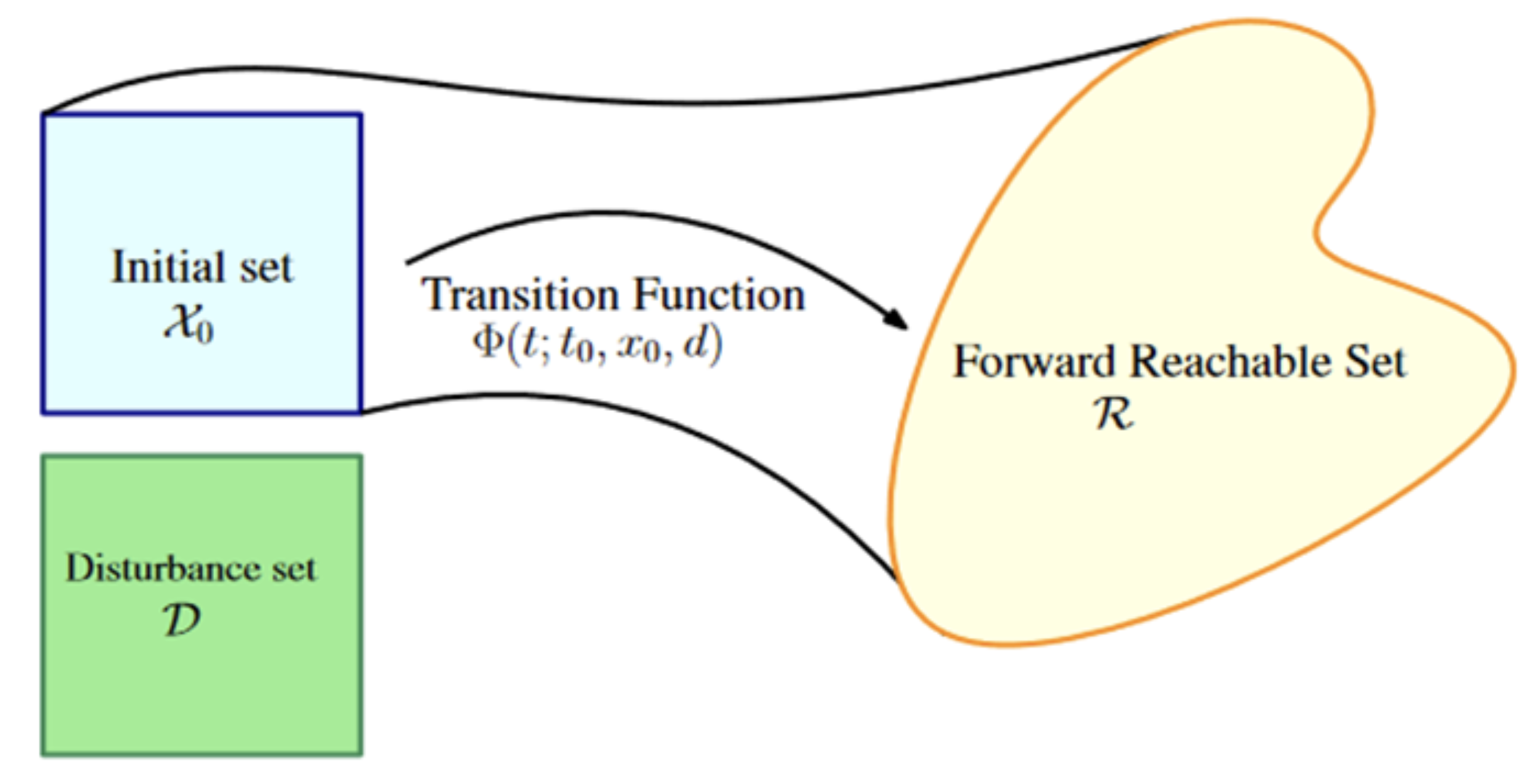}
    \caption{From \citep{devonport2021estimate}, an overview of data-driven reachability analysis.}
    \label{fig:my_label}
\end{figure}
\\
As long as the behavior of the system can be simulated or measured experimentally, it can be treated as a black-box model within the context of data-driven reachability analysis. 
\quad Rather than obtaining formal guarantees of the reachable set, as traditional approaches to reachability analysis do, data-driven approaches yield probabilistic guarantees of the estimates of the reachable set.
Samples $r_i = \Phi(t_1; t_0, x_{0i}, d_i)$, for $i = 1, \dots, N$ are drawn, where $x_{01}, \dots, x_{0N} \stackrel{i.i.d.}{\sim} X_0$ and $d_1, \dots, d_N \stackrel{i.i.d.}{\sim}D$, with random variables $X_0$ and $D$ defined on $\mathcal{X}_0$ and $\mathcal{D}$, respectively.
Let $C \subseteq 2^{\mathbb{R}^{n_x}}$ denote the class of admissible set estimators and $P_R$ the probability measure with respect to $R$. 
Then, a compact estimate $\hat{\mathcal{R}}$ of the reachable set is computed such that 
\begin{equation}\label{equation_1}
    P^N_R(P_R(\hat{\mathcal{R}}) \geq 1 - \epsilon) \geq 1 - \delta
\end{equation}
where $P^N_R$ is the product measure of $N$ copies of $P_R$, $\epsilon \in (0, 1)$ is the accuracy parameter, and $\delta \in (0, 1)$ is the confidence parameter \citep{devonport2021estimate}. 
\quad The double inequality in (\ref{equation_1}), a special case of the bound used in the Probably Approximately Correct (PAC) framework of statistical learning theory, provides two assertions. 
First, the inner inequality $P_R(\hat{\mathcal{R}}) \geq 1 - \epsilon$ asserts that $\hat{\mathcal{R}}$ attains a probability mass of at least $1 - \epsilon$ under $P_R$. 
Second, the outer inequality asserts that $\hat{\mathcal{R}}$ attains a $1 - \epsilon$ accuracy with probability $1 - \delta$ with respect to the samples $r_1, \dots, r_N$.
\subsection{Methods of Estimation}
\quad The DaDRA library incorporates two data-driven methods: a scenario approach to chance-constrained optimization with $p$-norm balls and an empirical risk minimization approach using a class of polynomials called empirical inverse Christoffel functions.
Both approaches are accompanied by a known lower bound for the number of samples $N$ in order to satisfy the specified probabilistic parameters $\epsilon$ and $\delta$.
\subsubsection{Scenario Reachability with \emph{p}-Norm Balls}
\quad The method of estimating reachable sets using $p$-norm balls is a scenario approach to chance-constrained convex optimization. 
For scenario reachability, we sample from our initial set $X_0$ and disturbance set $D$, and apply the state transition function $\Phi$ to compute samples $r_1, \dots, r_N \stackrel{i.i.d.}{\sim} R$.
We then find the set of parameters $\theta \in \Theta$, where $\Theta \subseteq \mathbb{R}^{n_\theta}$ is convex and compact, and the corresponding set $\hat{\mathcal{R}}(\theta)$ that minimizes some volume proxy subject to the constraint that $r_1, \dots, r_N \in \hat{\mathcal{R}}$.
Furthermore, the volume proxy is a function of the parameters, Vol$(\theta)$.
\quad In the case of $p$-norm balls, the reachable set estimate is
\begin{equation}
    \hat{\mathcal{R}}(A, b) = \{x \in \mathbb{R}^{n_x} : \lVert Ax - v \rVert_p \leq 1\},   
\end{equation}
where $A = A^\top \in \Theta_A \subseteq \mathbb{R}^{n_x \times n_x}$, $b \in \Theta_b \subseteq \mathbb{R}^{n_x}$, with compact $\Theta_A$, $\Theta_B$, and $\lVert\cdot\rVert_p$ denoting the $p$-norm.
The volume proxy that is minimized in this case is Vol$(A, b) = - \log\det A$, subject to the constraint $\lVert Ar_i - b \rVert \leq 1$, for all $i = 1, \dots, N$ \citep{devonport2021estimate}.
\quad For state dimension $n_x$, the number of samples required to meet the probabilistic guarantees with the algorithm using $p$-norm balls is
\begin{equation}
    N = \bigg\lceil\frac{1}{\epsilon}\frac{e}{e-1}\bigg(\log\frac{1}{\delta} + \frac{1}{2}(n^2_x + 3n_x)\bigg)\bigg\rceil
\end{equation}
\citep{devonport2021estimate}.

\subsubsection{Empirical Inverse Christoffel Function Method}
Given a finite measure $\mu$ on $R^n$ and a positive integer $k$, the empirical inverse Christoffel function is a polynomial of degree $2k$ defined by
\begin{equation}
\begin{split}
C(x) &= z_k(x)^\top \hat{M}^{-1}z_k(x) \\
&= z_k(x)^\top \bigg(\frac{1}{N}\sum_{i=1}^N z_k(x^{(i)})z_k(x^{(i)})^\top \bigg)^{-1}z_k(x)
\end{split} 
\end{equation}
where $z_k(x)$ is the vector of monomials of degree $\leq k$ and $\hat{M}$ is constructed from a collection of iid samples $x^{(i)}, i = 1, \dots, N$ from the probability distribution which is being estimated.
The empirical inverse Christoffel function method computes the $\alpha$-sublevel set of $C(x)$, that is $\{x \in \mathbb{R}^n : C(x) \leq \alpha \}$, to estimate the reachable set using iid samples $x_f^{(i)} = \Phi(t_1 ; t_0, x_0^{(i)}, d^{(i)})$, $\forall i \in \{1, \dots, N\}$, with $x_0^{(i)}$ from $X_0$ and $d^{(i)}$ from $D$, where $X_0$ and $D$ are defined on $\mathcal{X}_0$ and $\mathcal{D}$, respectively \citep{devonport2021datadriven}.

\quad For state dimension $n_x$, the number of samples required to meet the probabilistic guarantees with the algorithm using the empirical inverse Christoffel function method is
\begin{equation}
    N = \bigg\lceil\frac{5}{\epsilon}\bigg(\log\frac{4}{\delta} + \binom{n_x+2k}{n_x}\log\frac{40}{\epsilon}\bigg)\bigg\rceil
\end{equation}
\citep{devonport2021datadriven}.
\section{Features}
\quad The DaDRA library is built for ease of use while allowing the user enough autonomy to make specifications particular to the problem at hand.
Figure \ref{fig:hierarchy} illustrates an overview of the DaDRA library.
The three main modules are the \texttt{disturbance} module, the \texttt{dyn\_sys} module, and the \texttt{estimate} module.
The following sections will describe these modules in further detail.

\tikzstyle{startstop} = [circle, minimum width=1.5cm, minimum height=1cm,text centered, draw=black, fill=red!30]
\tikzstyle{interface} = [ellipse, minimum width=3cm, minimum height=1cm, text centered, draw=black, fill=orange!30]
\tikzstyle{class} = [rectangle, minimum width=2cm, minimum height=1cm, text centered, draw=black, fill=yellow!50]
\tikzstyle{module} = [rectangle, rounded corners, minimum width=2cm, minimum height=1cm, text centered, draw=black, fill=blue!20]
\tikzstyle{arrow} = [thick,->,>=stealth]
\tikzstyle{darrow} = [dashed, thick,->,>=stealth]

\begin{figure}[ht]
    \centering
    \begin{tikzpicture}[node distance=1.5cm]
        \node (dadra) [startstop] {DaDRA};
        \node (dynsys) [module, below of=dadra, yshift=-1cm] {\small \texttt{dyn\_sys} module};
        \node (disturbance) [module, left of=dynsys, xshift=-2cm] {\small \texttt{disturbance} module};
        \node (estimate) [module, right of=dynsys, xshift=2cm] {\small \texttt{estimate} module};
        \draw [arrow] (dadra) -- (disturbance);
        \draw [arrow] (dadra) -- (dynsys);
        \draw [arrow] (dadra) -- (estimate);
        \draw [darrow] (disturbance) -- (dynsys);
        \draw [darrow] (dynsys) -- (estimate);
        \node (system) [interface, below of=dynsys, yshift=-0.5cm] {\small \texttt{System} interface};
        \node (dsystem) [class, below of=system, yshift=-0.5cm] {\scriptsize \texttt{DisturbedSystem} class};
        \node (ssystem) [class, left of=dsystem, xshift=-1.5cm] {\scriptsize \texttt{SimpleSystem} class};
        \node (sampler) [class, right of=dsystem, xshift=1.5cm] {\scriptsize \texttt{Sampler} class};
        \draw [arrow] (dynsys) -- (system);
        \draw [arrow] (system) -- (dsystem);
        \draw [arrow] (system) -- (ssystem);
        \draw [arrow] (system) -- (sampler);
        \node (sdisturbance) [class, below of=disturbance, yshift=-0.5cm, xshift=-1cm] {\small \texttt{ScalarDisturbance} class};
        \node (cdisturbance) [class, below of=sdisturbance, xshift=-1.5cm, yshift=-0.5cm] {\scriptsize \texttt{Disturbance} class};
        \draw [arrow] (disturbance) -- (sdisturbance);
        \draw [darrow] (sdisturbance) -- (cdisturbance);
        \node (estimator) [class, below of=estimate, xshift=1cm, yshift=-0.5cm] {\small \texttt{Estimator} class};
        \draw [arrow] (estimate) -- (estimator);
    \end{tikzpicture}
    \caption{An overview of the DaDRA library.}
    \label{fig:hierarchy}
\end{figure}
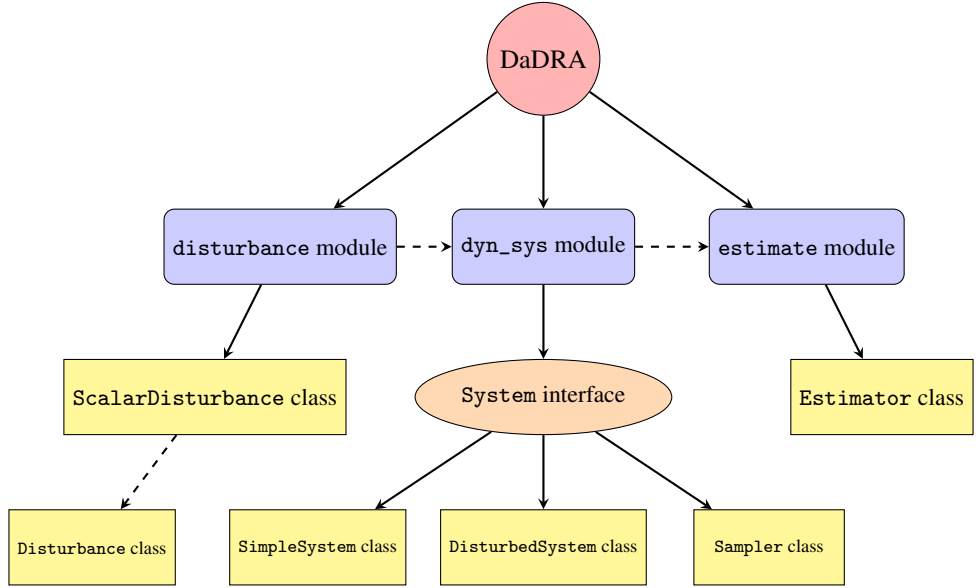
\subsection{The \texttt{disturbance} module}
As shown in Figure \ref{fig:hierarchy}, the \texttt{disturbance} module consists mainly of the \texttt{ScalarDisturbance} class and the \texttt{Disturbance} class. 
\subsubsection{The \texttt{ScalarDisturbance} class}
\quad The \texttt{ScalarDisturbance} class models the disturbance of a dynamic system along a single dimension or variable.
Within this class, the disturbance $d$ is a function of time and is modeled as a weighted sum of basis functions
\begin{equation}
    d(t, \underline{\alpha}) = \sum_{i=0}^{m}\alpha_if_i(t)
\end{equation}
where $t$ is the time at which the disturbance is observed, $\underline{\alpha}$ is the $m$-dimensional vector of weights, and $f_0, \dots, f_m$ are the basis functions, which themselves are each a function of time.
\quad The vector of weights $\underline{\alpha}$ is an $m$-dimensional random variable such that $\underline{\alpha} \sim Uniform([0, \frac{1}{i}]^m)$. 
The vector $\underline{\alpha}$ is initialized at the same time that the initial set $x_0 \in X_0$ is drawn, that is, at the beginning of each trajectory of the system.
For a \texttt{ScalarDisturbance} object, this is done using the \texttt{ScalarDisturbance.draw\_alpha()} instance method.
To obtain the disturbance for a given variable at a specific time, the \texttt{ScalarDisturbance.d(t)} method is used.

\quad Note that number of basis functions used does not affect the number of samples required from the system to satisfy the probabilistic parameters $\epsilon$ and $\delta$, as the disturbances only affect the random variable distribution of the system, rather than the dimensions of the system.
\quad An example of a set of $m$ basis functions is $\{f_i(t) : 1 \leq i \leq m\}$ such that
\begin{equation}
    f_i(t) = 
    \begin{cases}
    1 & i = 0 \\
    \sin(2\pi it) & i > 0
    \end{cases}
\end{equation}
and an instance of the \texttt{ScalarDisturbance} class with such a disturbance can automatically be created with the class method \texttt{ScalarDisturbance.sin\_disturbance()}.
\subsubsection{The \texttt{Disturbance} class}\label{cdisturbance}
\quad The \texttt{Disturbance} class extends the functionality of \texttt{ScalarDisturbance} for all $n$-dimensions of a system. In particular, an instance of \texttt{Disturbance} contains an instance of \texttt{ScalarDisturbance} for each variable. Furthermore, the \texttt{Disturbance.draw\_alphas()} instance method initializes the weights of the basis functions for the disturbances of each variable independently and the \texttt{Disturbance.get\_dist(n, t)} allows the disturbance of the $n$-th variable at time $t$ to be obtained.
\subsection{The \texttt{dyn\_sys} module}
\quad The \texttt{dyn\_sys} module provides the user with tools to model and sample from their system. The \texttt{dyn\_sys.System} interface defines a blueprint for a set of classes \texttt{dyn\_sys.SimpleSystem}, \texttt{dyn\_sys.DisturbedSystem}, \texttt{dyn\_sys.Sampler}, which each include a function defining the dynamics of the system, the degrees of freedom of the system, a set of intervals from which the initial states of the variables are drawn, and a means of sampling the system. Because data-driven reachability analysis requires iid samples from a system, these samples can be drawn in parallel using the classes that extend the \texttt{dyn\_sys.System} interface to speed up the process of sampling. 
\subsubsection{The \texttt{SimpleSystem} class}
\quad The \texttt{SimpleSystem} class provides a barebones implementation of a non-disturbed dynamic system.
The user need only specify the system dynamics and the intervals from which the initial state variables are drawn in order to create an object from which $N$ iid samples can be drawn using the instance method \texttt{SimpleSystem.sample\_system}.
\subsubsection{The \texttt{DisturbedSystem} class}
\quad The \texttt{DisturbedSystem} class makes use of the \texttt{Disturbance} class from section \ref{cdisturbance} to model a disturbed dynamic system.
Similar to the \texttt{SimpleSystem} class, the user specifies the system dynamics and the intervals from which the initial state variables are drawn in addition to the type of disturbance for the system.
\subsubsection{The \texttt{Sampler} class}\label{sampler}
\quad The \texttt{Sampler} class provides the user with greater autonomy over the specification of the system.
Note that the \texttt{SimpleSystem} and \texttt{DisturbedSystem} classes both limit the variables of the initial set to uniform random variables over intervals.
In contrast, the \texttt{Sampler} class acts as a wrapper, prompting the user to specify a means of sampling the system of interest. 
The details of the system, such as the random variables of the initial states and the disturbances for each variable, are implicit to the means of sampling the system. 
This class allows the user the ability to specify their system in the case that \texttt{SimpleSystem} and \texttt{DisturbedSystem} are too limiting, while maintaining the ability of the \texttt{Estimator} class to make use of the consistent properties of the \texttt{System} interface.
\subsection{The \texttt{estimate} module}
\subsubsection{The \texttt{Estimator} class}\label{estimator}
\quad The \texttt{estimate.Estimator} class allows the user to perform the actual data-driven reachability analysis process.
The user specifies their system using the classes from the \texttt{disturbance} and \texttt{dyn\_sys} modules.
A variety of class methods within the \texttt{Estimator} class provide different options for how to create an instance of the class, each offering their own advantages depending on the specifications of the user. 
\quad Upon instantiating an \texttt{Estimator} object, the user may specify or use the defualt values of a variety of parameters. 
These include the probabilistic parameters $\epsilon$ and $\delta$, the method of estimation (i.e. $p$-norm or Christoffel method), and various parameters pertaining to the method of estimation, for instance, the order of the Christoffel method or other constants. 
\quad In addition, the user can specify particular variables over which the analysis is to be computed over at the time of instantiating. 
Alternatively, they can use the \texttt{Estimator.iso\_dim(vars)} function after instantiating their object. 
Because the number of samples required depends on the state dimension and the number of variables being considered, by using \texttt{Estimator.iso\_dim(vars)}, the user can reduce the required number of samples in comparison with analysis of the full state system, while still using any variables from the full state system which the isolated variables may depend on.
\quad After initializing an instance of \texttt{Estimator}, the \texttt{Estimator.summary()} instance method allows the user to view information about the object.
This includes the state dimension, the accuracy and confidence parameter values, the number of samples required to satisfy those specified probabilistic parameter values, the parameters particular to the method of estimation, and whether or not a reachable set estimate has been computed yet.
\quad When the user calls \texttt{Estimator.sample\_system()} the required number of samples to satisfy the specified probabilistic parameter values is drawn from the system. 
These samples are drawn in parallel if the \texttt{Estimator} object was instantiated using one of the classes with the \texttt{dyn\_sys.System} interface.
The user can then use \texttt{Estimator.plot\_samples}, perhaps along with \texttt{Estimator.iso\_dim(vars)}, to visualize the trajectories of the system across up to three dimensions in addition to time.
\quad After sampling from the system, \texttt{Estimator.compute\_estimate()} computes the reachable set estimate based on the sample trajectories using the specified method of estimation. %
After computing the reachable set estimate, the user can visualize the set along with the trajectories using \texttt{Estimator.plot\_reachable()} or \texttt{Estimator.plot\_reachable\_time()}, yielding 2D or 3D plots with the optional of creating a gif that shows the progression of the system over time.
\quad Appendix \ref{chaoscode} shows the code for example usage of the DaDRA library for data-driven reachability analysis on the chaotic system described in section \ref{chaoticex}. 
Note the relatively small amount of code required using DaDRA to achieve high probabilistic guarantees on a system that is analytically intractable using most traditional methods of reachability analysis.
\section{Example Usage}

%
%
\subsection{Chaotic System Example}\label{chaoticex}
\quad To demonstrate the utility of the DaDRA library we perform reachability analysis on a Duffing oscillator, a kind of chaotic system. While traditional approaches tend to find difficulty performing reachability analysis on chaotic systems due to their analytically intractable nature, we can do so easily using DaDRA as a result of its data-driven approach. All we need to do is to draw samples from the system, and we gain enough information to compute a reachable set.

\quad The Duffing oscillator is a time-varying nonlinear oscillator with dynamics
\begin{equation}
\begin{split}
    \dot{x} &= y \\
    \dot{y} &= -\alpha y + x - x^3 + \gamma \cos{\omega t}
\end{split}
\end{equation}
with states $x, y \in \mathbb{R}$ and parameters $\alpha, \gamma, \omega \in \mathbb{R}$. In accordance with examples from \citet{devonport2021datadriven}, we choose values $\alpha = 0.05$, $\gamma = 0.4$, and $\omega = 1.3$, as the Duffing oscillator exhibits chaotic behavior for such values. In addition, the initial set is $(x(0), y(0)) \in X_0$, where $X_0$ is the uniform random variable over $([0.95, 1.05], [-0.05, 0.05])$. The time range is $[t_0, t_1] = [0, 100]$ and the state of the system $(x(t), y(t))$ for $t = 100$ is recorded for each sample.

\begin{figure}[ht]
    \centering
    \captionsetup{width=.8\linewidth}
    \includegraphics[width=.4\linewidth]{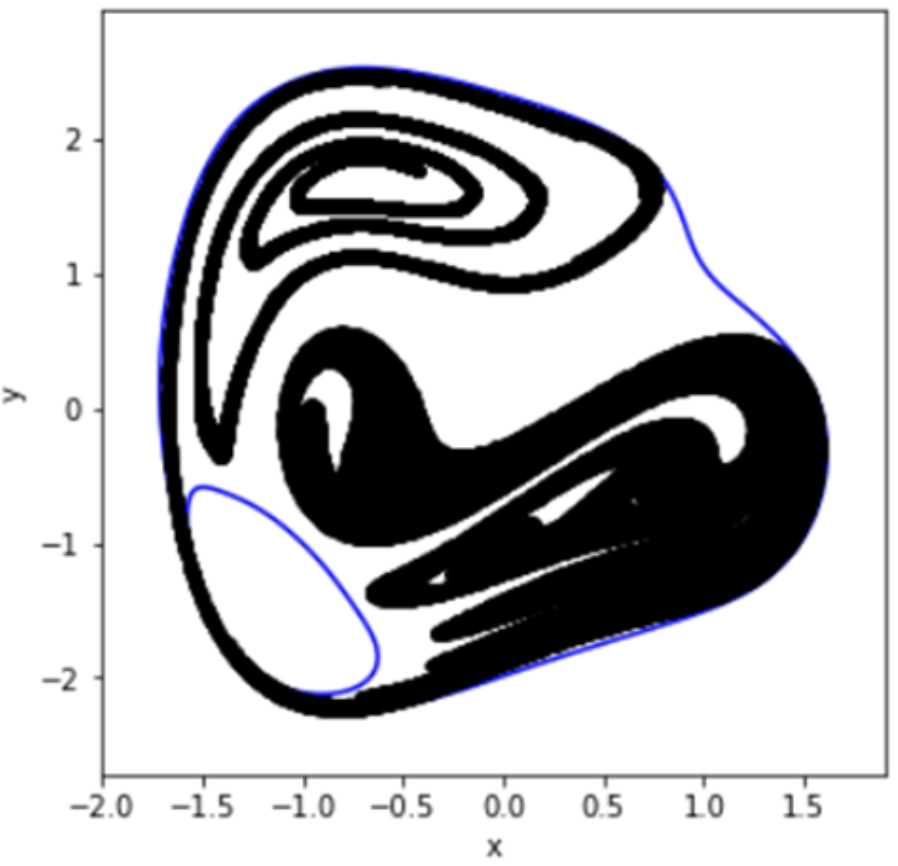}
    \caption{The reachable set estimate of a Duffing oscillator, computed and visualized using DaDRA.}
    \label{fig:chaotic_osci}
\end{figure}
\quad Figure \ref{fig:chaotic_osci} shows the samples from the Duffing oscillator in black, plotted along with the reachable set estimate, computed by DaDRA, in blue. 
As we can see, the reachable set estimate contains entirely the evolutions of the chaotic system with a tight bound. %
Notably, the area in the bottom left of the plot which contains none of the samples is correctly analyzed as an area which the chaotic system does not reach\textemdash this is illustrated by the hole in the blue reachable set estimate. In this case, the Christoffel function method was used with probabilistic parameters $\epsilon = 0.05$ and $\delta = 10{-9}$, corresponding to a reachable set estimate made with an expected 95\% accuracy and a 1 in a billion chance of failure. The code for this example is included in Appendix \ref{chaoscode}.
\subsection{Quadrotor Demonstration}
\subsubsection{Quadrotor Benchmark}\label{quadbench}
\quad In accordance with the quadrotor benchmark from \citep{arch19}, we compare the results of the data-driven methods built into DaDRA with the results of previous tools using traditional approaches to reachability analysis.
The 12-state quadrotor benchmark is meant to check control specifications for stabilization using PD controllers for height, roll, and pitch.
The objective is to control the quadrotor to change the height from 0 [m] to 1 [m] within 5 [s], reaching and the goal region of height $[0.98, 1.02]$ within $5$ [s] and remaining below $1.4$ for all times. 
After $1$ [s] the height should stay above $0.9$ [m]. 
The initial position and velocity of the quadrotor is uncertain in all directions within $[-0.4, 0.4]$ [m] and $[-0.4, 0.4]$ [m \slash s], respectively \citep{arch19}.
\begin{figure}[ht]
    \centering
    \captionsetup{width=.8\linewidth}
    \includegraphics[width=.8\linewidth]{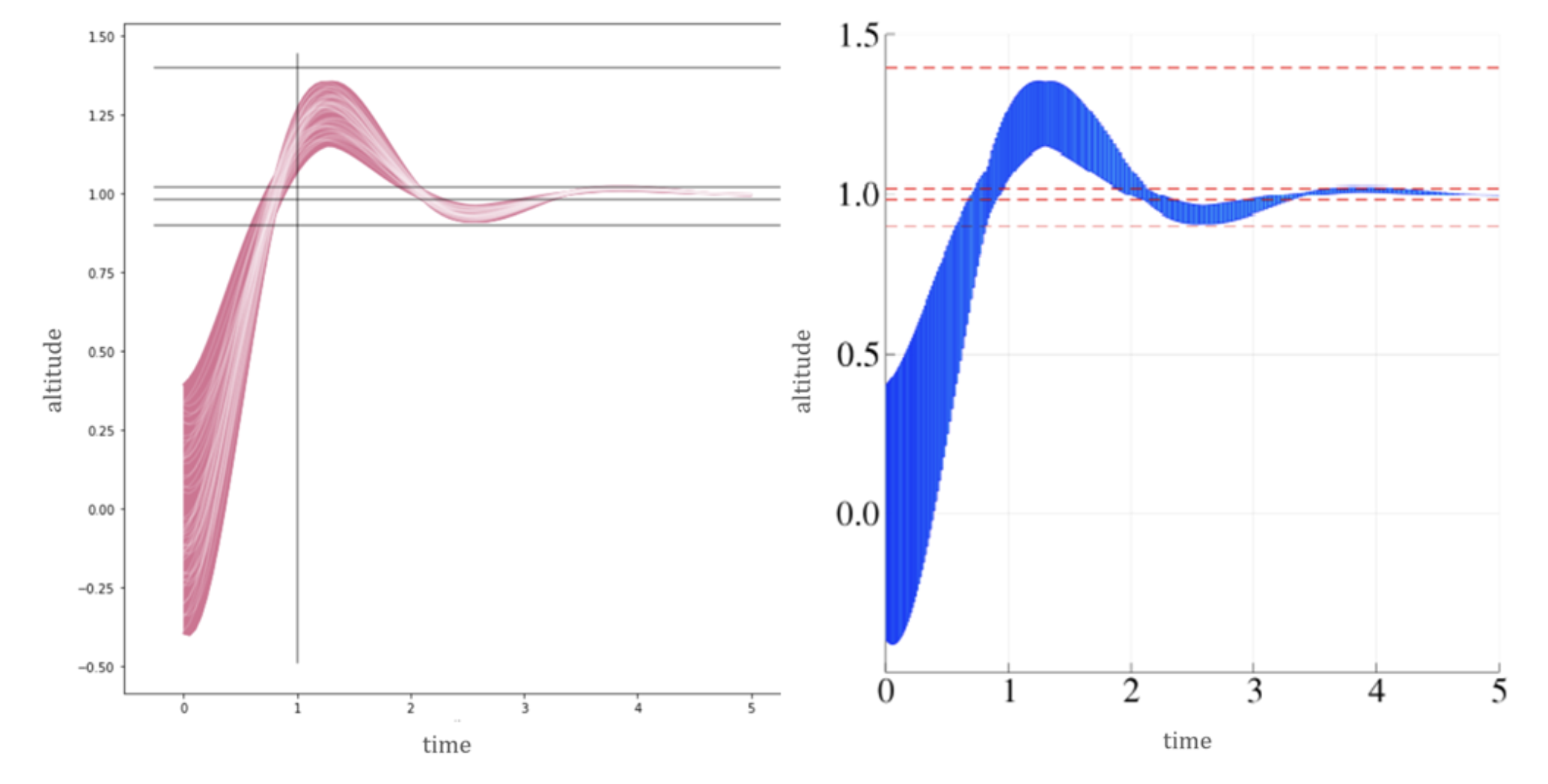}
    \caption{(left) The reachable set estimate computed and visualized by DaDRA in dark pink and the sample trajectories of the quadrotor illustrated by the light pink contours. (right) The reachable set estimate computed and visualized by JuliaReach in blue.}
    \label{fig:quad_bench}
\end{figure}
\quad Figure \ref{fig:quad_bench} shows the reachable set estimate across time for a single dimension, the altitude, computed and visualized by DaDRA, on the left, in pink, and the reachable set of the system computed and visualized by JuliaReach \citep{juliareach2019}, another tool using traditional methods for reachability analysis, on the right, in blue.
As we can see, the results are similar, serving as evidence that the data-driven methods of DaDRA are as effective as the traditional methods when applied to simplified systems.
\quad Further details about the quadrotor benchmark are described in Appendix \ref{quadbenchspec}. Appendix \ref{benchmarkperf} includes other benchmarks and a comparison of DaDRA performance with additional tools using traditional approaches to reachability analysis.
\quad Note that while the traditional methods employed by JuliaReach yield a formal guarantee of the reachable set, the data-driven methods of DaDRA provide a probabilstic guarantee with high confidence.
However, the benchmark quadrotor consists only of an undisturbed system with a simple controller for the purpose of satisfying the requirements of a goal region along just a single dimension.
Such a system would be unlikely to have any practical use in reality.
\subsubsection{Controller Synthesis for a Disturbed Quadrotor}
\begin{figure}[ht]
    \centering
    \captionsetup{width=.8\linewidth}
    \includegraphics[width=.8\linewidth]{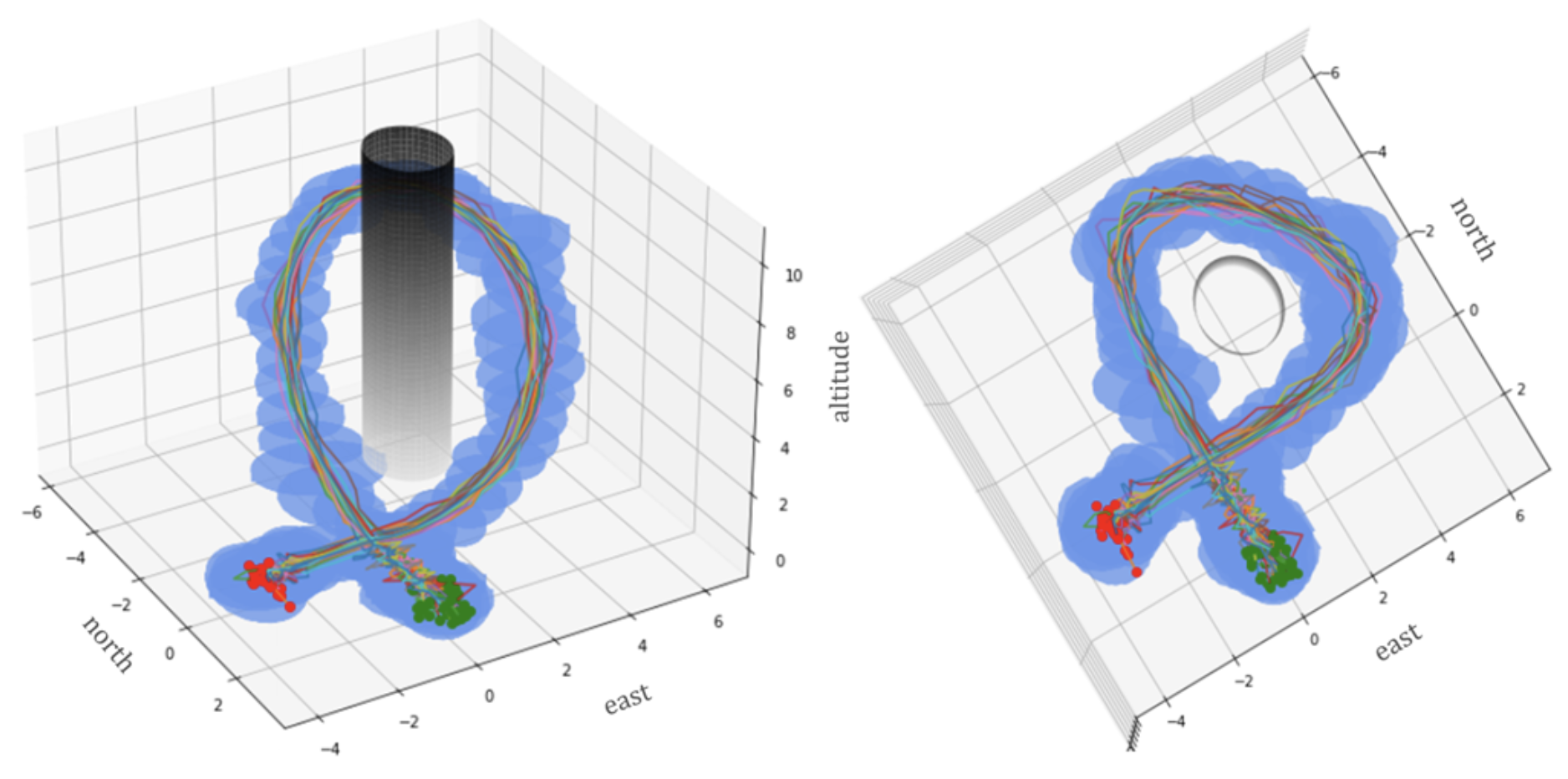}
    \caption{The reachable set estimate across time (in blue) of the 12-state quadrotor in 3 dimensions (north, east, and altitude), controlled to perform a clover-leaf maneuver while avoiding the unsafe set (the black cylinder). The green points denote the starting points for each of the rainbow colored sample trajectories, and the red points denote the end points.}
    \label{fig:dist_quad}
\end{figure}

\quad A much more insightful display of a tool for safety verification and controller design would necessitate analysis of a realistic system, including unpredictable disturbances and complex maneuvers. 
For example, as opposed to the simplified benchmark quadrotor system, a system in which the quadrotor has to satisfy some objective in three dimensions while remaining outside of an unsafe set and being acted upon by disturbance signals. 
\quad To demonstrate the full effectiveness of DaDRA, we perform reachability analysis on a 12-state quadrotor with added disturbance by a modified version of a military-specified wind turbulence model \citep{hakim2018} in order to tune a complex controller to perform a clover-leaf maneuver \citep{hall2012} in three dimensions while avoiding an unsafe cylindrical region. 
Applying DaDRA, we iteratively tuned a controller, using \citep{beard08quad} as a point of reference. 
The controller was tuned using the \texttt{dyn\_sys.Sampler} and \texttt{estimate.Estimator} classes described in sections \ref{sampler} and \ref{estimator}, respectively.
\quad The tuning process was as follows.
We first devised a controller based on \citep{beard08quad} to perform the clover-leaf maneuver in a system without disturbance.
We used the visualizations of the library to compare the sample trajectories of the system with the clover-leaf maneuver. We tuned the controller and compared the sample trajectories using DaDRA until the trajectories successfully followed a clover-leaf maneuver.
We then computed a reachable set estimate to determine whether the system reached the unsafe set, illustrated by the black cylinder in Figure \ref{fig:dist_quad}.
We modified the controller again until the reachable set estimate, the blue region in Figure \ref{fig:dist_quad}, successfully followed the clover-leaf maneuver while avoiding the unsafe set.
Finally, we added a modified version of the disturbance specified in \citep{hakim2018} and increased the magnitude of the disturbance until the reachable set intersected with the safe set, and then we chose the most extreme parameter values of the complex disturbance prior to our reachable set estimate intersecting with the unsafe set.
\quad Hence, the result was a tuned controller for a 12-state controller to perform a clover-leaf maneuver, as well as the most extreme disturbance conditions for which the quadrotor could perform the maneuver without reaching the unsafe set.
\quad Figure \ref{fig:dist_quad} shows the visualization provided by DaDRA of the reachable set estimate (in blue) of the disturbed quadrotor in 3 dimensions (north, east, and altitude). The reachable set estimate was computed with probabilistic parameters $\epsilon = 0.05$ and $\delta = 10^{-9}$, corresponding to a $95\%$ accuracy with a 1 in a billion chance of failure. As we can see, the reachable set estimate does not intersect with the unsafe set (the black cylindrical region), and hence, the library allowed us to successfully iteratively tune a non-trivial controller for a system acted upon by a complex disturbance while avoiding an unsafe region with near certainty.

\section{Conclusion}
\quad Though traditional approaches to reachability analysis have the advantage of providing formal guarantees of the reachable set, they tend to be impractical for use in real world settings, as most systems of interest possess high-dimensional, analytically intractable, and possibly unknown dynamics.
Applying the data-driven methods allows for reachability analysis with arbitrarily robust probabilistic guarantees, so long as the system of interest is capable of being simulated or sampled from.
\quad DaDRA takes advantage of data-driven methods in order to provide an easy-to-use alternative to libraries implementing traditional algorithms for reachability analysis. We demonstrate the practical functionality of DaDRA initially on a chaotic system and subsequently on a realistic system acting under complex disturbance signals and controlled with an intricate controller across multiple dimensions. The examples outline the utility of the library on analytically intractable systems, particularly for the purpose of safety verification and controller design.
\section*{Acknowledgments}
Jared Mejia would like to thank Professor Arcak Murat and PhD student Alex Devonport for their excellent mentorship and guidance. 
Thank you to Leslie Mach and the UC Berkeley EECS department for organizing the 2021 SUPERB REU Program, and thank you to the NSF for funding this program.

\bibliographystyle{unsrtnat}
\bibliography{references}

\appendix
\appendixpage
\section{Benchmark Performance}\label{benchmarkperf}

\subsection{Laub-Loomis Benchmark}

\quad The dynamics for the Laub-Loomis model \citep{loomis1998} is defined by an ODE with 7 variables:

\begin{equation}
    \begin{cases}
      \dot{x}_1 = 1.4x_3 - 0.9x_1 \\
      \dot{x}_2 = 2.5x_5 - 1.5x_2 \\
      \dot{x}_3 = 0.6x_7 - 0.8x_2x_3 \\
      \dot{x}_4 = 2 - 1.3x_3x_4 \\
      \dot{x}_5 = 0.7x_1 - x_4x_5 \\
      \dot{x}_6 = 0.3x_1 - 3.1x_6 \\
      \dot{x}_7 = 1.8x_6 - 1.5x_2x_7
    \end{cases}
\end{equation}
The initial sets of the model are boxes centered at $x_1(0) = 1.2$, $x_2(0) = 1.05$, $x_3(0) = 1.5$, $x_4(0) = 2.4$, $x_5(0) = 1$, $x_6(0) = 0.1$, and $x_7(0) = 0.45$, and the range of the box for the $i$th dimension is defined by the interval $[x_i(0) - W, x_i(0) + W]$ \citep{romain2013nltoolbox}. 
Note that the larger the initial set, the harder the reachability analysis for traditional approaches to reachability analysis. The benchmark defined in \citep{arch19} considers width of the box $W = 0.1$ with an unsafe set defined by $x_4 \geq 5$ and a time horizon of $[0, 20]$.
\begin{figure}[ht]
    \centering
    \captionsetup{width=.8\linewidth}
    \includegraphics[width=.45\linewidth]{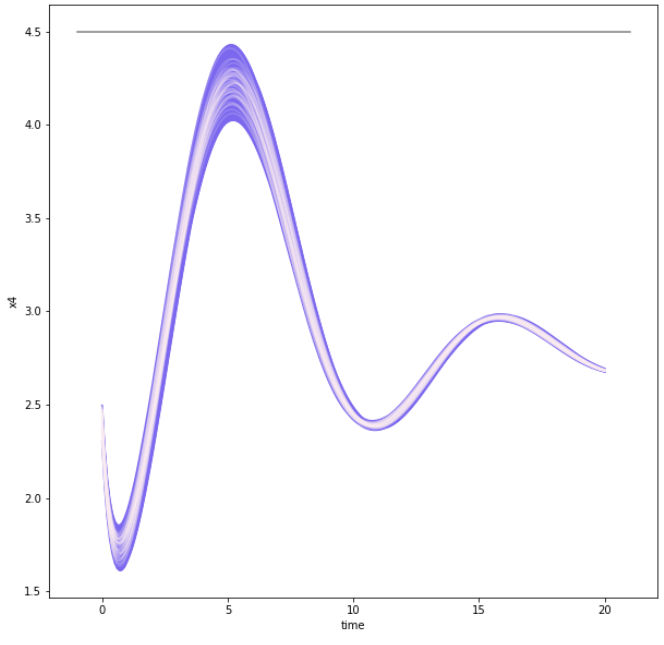}
    \includegraphics[width=.45\linewidth]{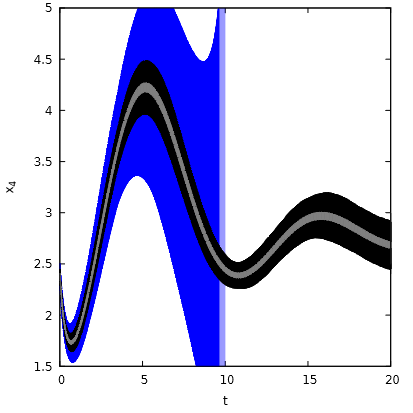}
    \caption{(left) The reachable set estimate of the Laub-Loomis benchmark computed and visualized by DaDRA for $W = 0.1$ in purple and the sample trajectories of the system illustrated by the light pink contours. (right) The reachable set estimate of the Laub-Loomis benchmark computed and visualized by Isabelle/HOLf for $W = 0.01$, $W = 0.05$, and $W = 0.1$ in grey, black, and blue, respectively.}
    \label{fig:laub-loom-dadra}
\end{figure}
\quad Figure \ref{fig:laub-loom-dadra} shows the reachable set estimate of the variable $x_4$ across time for both DaDRA and Isabelle/HOL \citep{isabelle2015hol}, another library using traditional approaches for reachability analysis.
As can be seen, though Isabelle/HOL succesfully computes relatively precise enclosures of the reachable set for $W = 0.01$ and $W = 0.05$, the tool over approximates the reachable set for $W = 0.01$ and fails entirely to maintain reasonable enclosures for $t \geq 7$. 
In comparison, the reachable set estimate made by DaDRA is very precise and still includes all of the trajectories of the system.
\subsection{Space Rendezvous Benchmark}
\quad The nonlinear dynamic equations and specifications from \citep{chan2017arch} describe the two-dimensional, planar motion of a spacecraft on an orbital plane towards a space station:
\begin{equation}
    \begin{cases}
      \dot{x} = v_x \\
      \dot{y} = v_y \\
      \dot{v_x} = n^2x + 2nv_y + \frac{\mu}{r^2} - \frac{\mu}{r^3}(r + x) + \frac{\mu_x}{m_c} \\
      \dot{v_y} = n^2y - 2nv_x - \frac{\mu}{r^3_c}y + \frac{u_y}{m_c}
    \end{cases}
\end{equation}
\quad The system states are $\underline{x} =$ ($x$ $y$ $v_x$ $v_y$)$^T$ with position relative to the target $x$, $y$ [m], time $t$ [min], horizontal velocity $v_x$ [m / min], and vertical velocity $v_y$ [m / min].
The parameters are $\mu = 3.986 \times 10^14 \times 60^2$ [m$^3$ / min $^2$], $r = 42164 \times 10^3$ [m], $m_c = 500$ [kg], $n = \sqrt{\frac{\mu}{r^3}}$ and $r_c = \sqrt{(r + x)^2 + y^2}$. 
\quad The switched controller consist of modes \emph{approaching} ($x \in [-1000, -100]$ [m]), \emph{rendezvous attempt} ($x \geq -100$ [m]), and \emph{aborting} (time $t \geq 120$ [min]).
The linear feedback controllers are $\binom{u_x}{u_y} = K_1\underline{x}$ for \emph{approaching} mode, $\binom{u_x}{u_y} = K_2\underline{x}$ for \emph{rendezvous attempt} mode, and $\binom{u_x}{u_y} = \binom{0}{0}$ for \emph{aborting} mode. The feedback matrices $K_i$ are:
\begin{equation}
\begin{split}
    K_1 = 
    \begin{pmatrix}
      -28.8287 & 0.1005 & -1449.9754 & 0.0046 \\
      -0.087 & -33.2562 & 0.00462 & -1451.5013
    \end{pmatrix}
    \\
    K_2 = 
    \begin{pmatrix}
      -288.0288 & 0.1312 & -96149898 & 0 \\
      -0.1312 & -288 & 0 & -9614.9883
    \end{pmatrix}
\end{split}
\end{equation}
\quad The initial set of the spacecraft is $x \in [-925, -875]$ [m], $y \in [-425, -375]$ [m], $v_x = 0$ [m/min] and $v_y = 0$ [m/min] and the considered time horizon is $t \in [0, 200]$ [min].
\begin{figure}[ht]
    \centering
    \captionsetup{width=.8\linewidth}
    \includegraphics[width=.45\linewidth]{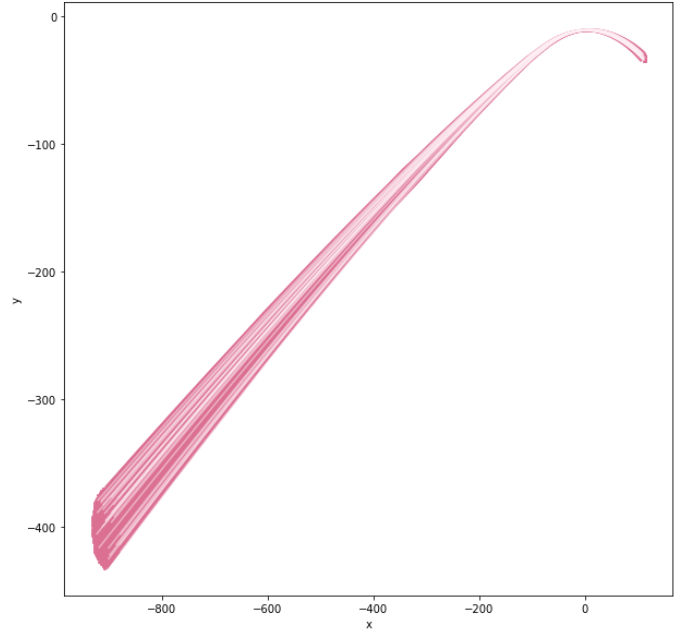}
    \includegraphics[width=.44\linewidth]{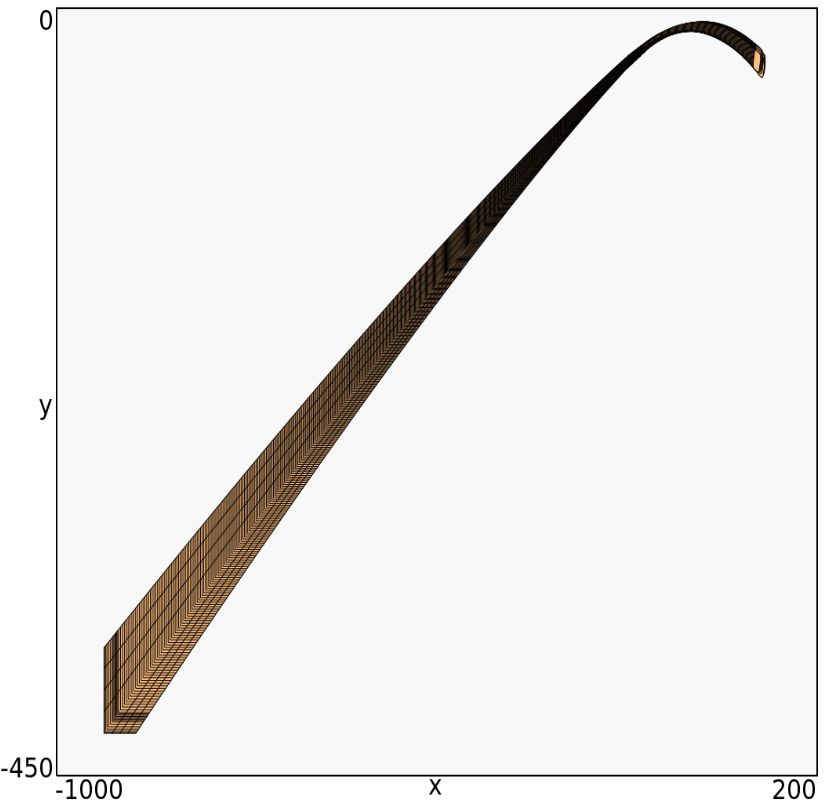}
    \caption{(left) The reachable set estimate of the Space Rendezvous benchmark computed and visualized by DaDRA dark pink and the sample trajectories of the system illustrated by the light pink contours. (right) The reachable set estimate of the Space Rendezvous benchmark computed and visualized by Ariadne.}
    \label{fig:spacer-dadra}
\end{figure}
\quad Figure \ref{fig:spacer-dadra} shows the reachable set estimate of the spacecraft in the $xy$-plane for the considered time horizon for both DaDRA and Ariadne \citep{balluchi2006ariadne}, another library using traditional approaches for reachability analysis. As can be seen, the computed reachable sets for both libraries appear similar.
\subsection{Quadrotor Benchmark Specification}\label{quadbenchspec}
\quad The following includes further details of the specification of the quadrotor benchmark as introduced in section \ref{quadbench} from \citep{arch19}. 
The variables of the model are the inertial (north) position $x_1$, the inertial (east) position $x_2$, the altitude $x_3$, the longitudinal velocity $x_4$, the lateral velocity $x_5$, the vertical velocity $x_6$, the roll angle $x_7$, the pitch angle $x_8$, the yaw angle $x_9$, the roll rate $x_10$, the pitch rate $x_11$, and the yaw rate $x_12$.
The required parameters are the gravity constant $g = 9.81$ [m / s$^2$], the radius of center mass $R = 0.1$ [m], the distance of motors to center mass $l = 0.5$ [m], motor mass $M_{rotor} = 0.1$ [kg], center mass $M = 1$ [kg], and total mass $m = M + 4M_{rotor}$.
\quad The moments of inertia are computed by
\begin{equation}
    \begin{split}
        J_x &= \frac{2}{5}MR^2 + 2l^2M_{rotor}, \\
        J_y &= J_x, \\
        J_z &= \frac{2}{5}MR^2 + 4l^2M_{rotor}.
    \end{split}
\end{equation}
\quad The set of ordinary differential equations for the quadrotor are
\begin{equation}
    \begin{cases}
      \dot{x}_1 &= \cos(x_8)\cos(x_9)x_4 + \big(\sin(x_7)\sin(x_8)\cos(x_9) - \cos(x_7)\sin(x_9)\big)x_5 \\&\textrm{\quad} + \big(\cos(x_7)\sin(x_8)\cos(x_9) + \sin(x_7)\sin(x_9)\big)x_6 \\
      \dot{x}_2 &= \cos(x_8)\cos(x_9)x_4 + \big(\sin(x_7)\sin(x_8)\sin(x_9) + \cos(x_7)\cos(x_9)\big)x_5 \\&\textrm{\quad} + \big(\cos(x_7)\sin(x_8)\sin(x_9) - \sin(x_7)\cos(x_9)\big)x_6 \\
      \dot{x}_3 &= \sin(x_8)x_4 - \sin(x_7)\cos(x_8)x_5 - \cos(x_7)\cos(x_8)x_6 \\
      \dot{x}_4 &= x_{12}x_5 - x_{11}x_6 - g\sin(x_8) \\
      \dot{x}_5 &= x_{10}x_6 - x_{12}x_4 + g\cos(x_8)\sin(x_7) \\
      \dot{x}_6 &= x_{11}x_4 - x_{10}x_5 + g\cos(x_8)\cos(x_7) - \frac{F}{m} \\
      \dot{x}_7 &= x_{10} + \sin(x_7)\tan(x_8)x_{11} + \cos(x_7)\tan(x_8)x_{12} \\
      \dot{x}_8 &= \cos(x_7)x_{11} - \sin(x_7)x_{12} \\
      \dot{x}_9 &= \frac{\sin(x_7)}{\cos(x_8)}x_11 + \frac{\cos(x_7)}{\cos(x_8)}x_12 \\
      \dot{x}_{10} &= \frac{J_y - J_z}{J_x}x_{11}x_{12} + \frac{1}{J_x}\tau_{\phi} \\
      \dot{x}_{11} &= \frac{J_z - J_x}{J_y}x_{10}x_{12} + \frac{1}{J_y}\tau_{\theta} \\
      \dot{x}_{12} &= \frac{J_x - J_y}{J_z}x_{10}x_{11} + \frac{1}{J_z}\tau_{\psi} 
    \end{cases}
\end{equation}
\quad The quadrotor is stabilized using PD controllers for height, roll, and pitch. The equations of the controllers are
\begin{align*}
        F &= mg - 10(x_3 - U_1) + 3x_6 & \textrm{(height control)}, \\
        \tau_{\psi} &= - (x_7 - u_2) - x_{10} & \textrm{(roll control)}, \\
        \tau_{\theta} &= -(x_8 - u_3) - x_{11} & \textrm{(pitch control)}, 
\end{align*}
where $u_1$, $u_2$, and $u_3$ are the desired values for height, roll, and pitch, respectively. The heading is left uncontrolled and so $\tau_{\psi} = 0$.
\quad The task of the quadrotor benchmark is described in section \ref{quadbench} and the results of the computed reachable sets is shown in Figure \ref{fig:quad_bench}.
\section{Chaotic System Analysis Code}\label{chaoscode}

The following is an example of using the DaDRA library corresponding to the analysis of the chaotic system described in section \ref{chaoticex}.

\definecolor{codegreen}{rgb}{0,0.6,0}
\definecolor{codegray}{rgb}{0.5,0.5,0.5}
\definecolor{codepurple}{rgb}{0.58,0,0.82}
\definecolor{backcolour}{rgb}{0.95,0.95,0.92}

\lstdefinestyle{mystyle}{
    backgroundcolor=\color{backcolour},   
    commentstyle=\color{codegreen},
    keywordstyle=\color{magenta},
    numberstyle=\tiny\color{codegray},
    stringstyle=\color{codepurple},
    basicstyle=\ttfamily\footnotesize,
    breakatwhitespace=false,         
    breaklines=true,                 
    captionpos=b,                    
    keepspaces=true,                 
    numbers=left,                    
    numbersep=5pt,                  
    showspaces=false,                
    showstringspaces=false,
    showtabs=false,                  
    tabsize=2,
    basicstyle=\scriptsize
}

\lstset{style=mystyle}

\begin{lstlisting}[language=Python]
import dadra
import numpy as np

# define the dynamics of the system
def duffing_oscillator(y, t, alpha=0.05, omega=1.3, gamma=0.4):
    dydt = [y[1], -alpha * y[1] + y[0] - y[0] ** 3 + gamma * np.cos(omega * t)]
    return dydt

# define the intervals for the initial states of the variables in the system
d_state_dim = 2
d_intervals = [(0.95, 1.05), (-0.05, 0.05)]

# instantiate a SimpleSystem object for a non-disturbed system
d_ds = dadra.SimpleSystem(
    dyn_func=duffing_oscillator,
    intervals=d_intervals,
    state_dim=d_state_dim,
    timesteps=100,
    parts=1001,
)

# instantiate an Estimator object
d_e = dadra.Estimator(
    dyn_sys=d_ds,
    epsilon=0.05,
    delta=1e-9,
    christoffel=True,
    normalize=True,
    d=10,
    rho=0.0001,
    rbf=False,
    scale=1.0,
)

# print out a summary of the Estimator object
d_e.summary()
"""
-----------------------------------------------------------------------
Estimator Summary
=======================================================================
State dimension: 2
Accuracy parameter epsilon: 0.05
Confidence parameter delta: 1e-09
Number of samples: 156626
Method of estimation: Inverse Christoffel Function
Degree of polynomial features: 10
Kernelized: False
Constant rho: 0.0001
Radical basis function kernel: False
Length scale of kernel: 1.0
Status of Christoffel function estimate: No estimate has been made yet
-----------------------------------------------------------------------
"""

# make a reachable set estimate on the disturbed system
d_e.estimate()
"""
Drawing 156626 samples
Using 16 CPUs

100%
156626/156626 [03:46<00:00, 683.80it/s]

Time to draw 156626 samples: 03 minutes and 47 seconds
Time to apply polynomial mapping to data: 00 minutes and 00 seconds
Time to construct moment matrix: 00 minutes and 00 seconds
Time to (pseudo)invert moment matrix: 00 minutes and 00 seconds
Time to compute level parameter: 00 minutes and 00 seconds
"""

# save a plot of the 2D contours of the estimated reachable set
d_e.plot_reachable("figures/d_estimate_2D.png", grid_n=200)
"""
Time to compute contour: 00 minutes and 00 seconds
"""
\end{lstlisting}
Note the relatively small amount of code required to perform the analysis using the DaDRA library.

\end{document}